\documentclass[aps,prd,preprintnumbers,superscriptaddress,nofootinbib,showpacs,twocolumn]{revtex4}%
\usepackage[dvipdfmx]{graphicx}
\usepackage{bm,latexsym,amsmath,amssymb,amsfonts,mathrsfs}
\usepackage{color}
\input{colordvi.tex}
\usepackage[dvipdfmx]{hyperref}
\usepackage{url}
\hypersetup{
    colorlinks=true,
    citecolor=cyan,
}
\newcommand*{\D}{{\rm d}}
\newcommand*{\mpl}{M_{\rm Pl}}
\begin{document}

\title{Generalized Galilean Genesis}

\author{Sakine~Nishi}
\email[Email: ]{sakine\_n"at"rikkyo.ac.jp}
\affiliation{Department of Physics, Rikkyo University, Toshima, Tokyo 175-8501, Japan
}

\author{Tsutomu~Kobayashi}
\email[Email: ]{tsutomu"at"rikkyo.ac.jp}
\affiliation{Department of Physics, Rikkyo University, Toshima, Tokyo 175-8501, Japan
}

\begin{abstract}
The galilean genesis scenario is an alternative to inflation
in which the universe starts expanding from Minkowski in the asymptotic past
by violating the null energy condition stably.
Several concrete models of galilean genesis
have been constructed so far within the context of galileon-type
scalar-field theories.
We give a generic, unified description of the galilean genesis scenario
in terms of the Horndeski theory, i.e., the most general scalar-tensor theory
with second-order field equations. In doing so we generalize the previous models
to have a new parameter (denoted by $\alpha$) which results in controlling the evolution of the Hubble rate.
The background dynamics is investigated to show that
the generalized galilean genesis solution is an attractor, similarly to the original model.
We also study the nature of primordial perturbations
in the generalized galilean genesis scenario.
In all the models described by our generalized genesis Lagrangian,
amplification of tensor perturbations does not occur as opposed to
what happens in quasi-de Sitter inflation.
We show that
the spectral index of curvature perturbations is
determined solely from the parameter $\alpha$ and does not depend on
the other details of the model.
In contrast to the original model,
a nearly scale-invariant spectrum of curvature perturbations is obtained
for a specific choice of $\alpha$.
\end{abstract}

\preprint{RUP-15-1}
\maketitle

\section{Introduction}

It is fair to say that inflation~\cite{r2, inflation1, inflation2}
followed by a hot Big Bang is a standard scenario of modern cosmology.
Inflation is attractive because the period of quasi-de Sitter expansion
in the early universe resolves several problems
that would otherwise indicate the need for fine-tuning.
Moreover, curvature perturbations are naturally generated from quantum fluctuations
during inflation, which seed large-scale structure of the universe~\cite{Mukhanov:1981xt}.
The basic prediction of inflation is that the primordial curvature perturbations
are nearly scale-invariant, adiabatic, and Gaussian.
This is in agreement with
observations of CMB anisotropies~\cite{Larson:2010gs, Ade:2013uln}.
Inflationary models also predict the quantum mechanical production of gravitational waves~\cite{Starobinsky:1980te},
the detection of which would be the evidence for inflation.

Despite the success of inflation, it would be reasonable to ask
whether only inflation can be a consistent scenario compatible with observations.
It should also be noted that an inflationary universe is
past geodesically incomplete~\cite{Borde:1996pt}
and so the problem of an initial singularity still persists.
From this viewpoint, various alternative scenarios have been proposed so far,
such as bouncing models.
Although such models can eliminate the initial singularity,
many of them are unfortunately plagued by
instabilities originated from the violation of the null energy condition (NEC),
the growth of shear, and primordial perturbations incompatible with observations~\cite{Battefeld:2014uga}.

In the context of cosmology, the violation of the NEC implies that
\begin{eqnarray}
\frac{\D H}{\D t}>0,
\end{eqnarray}
where $H$ is the Hubble rate and $t$ is cosmic time.
This signals ghost instabilities in general relativity.
Recently, however, it was noticed that
in noncanonical galileon-type scalar-field theories
the NEC can be violated {\em stably},\footnote{The NEC can be violated stability at least within linear perturbation analysis.  
However, at nonlinear order, it is not clear whether there are no instabilities \cite{Sawicki:2012uga}.} 
and based on this idea,
Creminelli {\em et al.} proposed a novel, stable alternative to inflation named
galilean genesis~\cite{Creminelli:2010ba}.
(See also Ref.~\cite{Creminelli:2006xe}.)
In the galilean genesis scenario, the universe is asymptotically Minkowski in the past and
starts expanding from this low energy state.
As such, this scenario is devoid of the horizon and flatness problems.
Aspects of galilean genesis have been studied in
Refs.~\cite{LevasseurPerreault:2011mw,Wang:2012bq,Rubakov:2013kaa,Elder:2013gya,Easson:2013ju} and
the original model has been extended in Refs.~\cite{Creminelli:2012my,Hinterbichler:2012fr,Hinterbichler:2012yn}
to possess improved properties.
See also
Refs.~\cite{Deffayet:2010qz,Kobayashi:2010cm,Qiu:2011cy,Easson:2011zy,Cai:2012va,Cai:2013vm,Osipov:2013ssa,Qiu:2013eoa,Pirtskhalava:2014oc}
for other interesting NEC violating cosmologies in galileon-type theories
and Ref.~\cite{Rubakov:2014jja} for a related review.

In this paper, we introduce a unified treatment of the galilean genesis models
and give a generic Lagrangian admitting the genesis solutions.
This is done by using the Horndeski theory~\cite{Horndeski},
which is the most general scalar-tensor theory with second-order field equations.
Our generalized galilean genesis Lagrangian contains four functional degrees of freedom and
a constant parameter denoted $\alpha$. This parameter determines the behavior of the Hubble rate.
For specific choices of those functions and $\alpha = 1$,
our Lagrangian reproduces the previous models explored in
Refs.~\cite{Creminelli:2010ba,Creminelli:2012my,Hinterbichler:2012fr,Hinterbichler:2012yn}.
As is often the case with inflation alternatives, it turns out that
the galilean genesis models in general fail to produce nearly scale-invariant curvature perturbations.
We show, however, that with an appropriate tuning of $\alpha$
it is possible to have a slightly tilted spectrum consistent with observations.

The Horndeski theory was developed about forty years ago~\cite{Horndeski}
and was revived recently as the generalized galileon theory~\cite{Deffayet:2011gz}.
The equivalence of the two theories was proven for the first time in Ref.~\cite{Kobayashi:2011nu}.
The action of the Horndeski theory is given
in the generalized galileon form by
\begin{eqnarray}
S&=&\int\D^4x\sqrt{-g}\left({\cal L}_2
+{\cal L}_3+{\cal L}_4+{\cal L}_5\right),\label{Hor}
\end{eqnarray}
with
\begin{eqnarray}
&&{\cal L}_2=G_2(\phi, X),\quad
{\cal L}_3=-G_3(\phi, X)\Box\phi,
\nonumber\\&&
{\cal L}_4=G_4(\phi, X)R
+G_{4X}\left[(\Box\phi)^2-(\nabla_\mu\nabla_\nu\phi)^2\right],
\nonumber\\&&
{\cal L}_5=G_5(\phi, X)G^{\mu\nu}\nabla_\mu\nabla_\nu\phi
-\frac{1}{6}G_{5X}\bigl[(\Box\phi)^3
\nonumber\\&&\qquad\qquad
-3\Box\phi(\nabla_\mu\nabla_\nu\phi)^2
+2(\nabla_\mu\nabla_\nu\phi)^3\bigr],
\end{eqnarray}
where $R$ is the Ricci scalar, $G_{\mu\nu}$ is the Einstein tensor, and each
$G_i(\phi, X)\;(i=2,3,4,5)$ is an arbitrary function of
the scalar field $\phi$ and $X:=-g^{\mu\nu}\partial_\mu\phi\partial_\nu\phi/2$.
We use the notation $G_{iX}$ to denote $\partial G_i/\partial X$.

The plan of this paper is as follows.
In the next section,
we present a generic Lagrangian that admits the generalized galilean genesis solution.
In Sec.~III we analyze the background evolution analytically and numerically
and show that the generalized galilean genesis solution is the dynamical attractor
for a wide range of initial conditions.
Primordial tensor and scalar perturbations from generalized galilean genesis
are studied in Sec.~IV, and
the curvaton mechanism in the genesis scenario is briefly discussed in Sec.~V.
We draw our conclusions in Sec.~VI.

\section{Generalized genesis solutions}

The original model of galilean genesis is
constructed by using the Lagrangian of the form~\cite{Creminelli:2010ba,Creminelli:2012my}
\begin{eqnarray}
{\cal L}=\frac{\mpl^2}{2}R+f_1 e^{2\lambda \phi} X+f_2 X^2+f_3 X\Box\phi,
\label{original-Lag}
\end{eqnarray}
where $f_1$, $f_2$, $f_3$, and $\lambda$ are constants.
(We have changed notations of Refs.~\cite{Creminelli:2010ba,Creminelli:2012my}.)
The above Lagrangian has the genesis solution,
\begin{eqnarray}
e^{\lambda\phi}\simeq \frac{{\rm const}}{-t},\quad
H\simeq \frac{{\rm const}}{(-t)^3}\quad (-\infty <t<0),\label{original-gen-sol}
\end{eqnarray}
for large $|t|$.
(We have a degree of freedom to shift the origin of time: $t\to t-t_0$.)
The scale factor is given by
$a\simeq 1+{\rm const}/(-t)^2$, describing the universe
that starts expanding from singularity-free Minkowski in the asymptotic past.
The same genesis solution can also be obtained from the DBI conformal
galileons~\cite{Hinterbichler:2012fr,Hinterbichler:2012yn}.

In Ref.~\cite{Nishi:2014} it was noticed that
the genesis solution~(\ref{original-gen-sol}) is obtained generically in
the subclass of the Horndeski theory with
\begin{eqnarray}
&&
G_2=e^{4\lambda\phi}g_2(Y),\quad G_3=e^{2\lambda\phi}g_3(Y),
\nonumber\\&&
G_4=\frac{\mpl^2}{2}+e^{2\lambda\phi}g_4(Y),
\quad
G_5=e^{-2\lambda\phi}g_5(Y),\label{gen-Horn}
\end{eqnarray}
where each $g_i$ ($i=2,3,4,5$) is an arbitrary function of
\begin{eqnarray}
Y:= e^{-2 \lambda\phi} X.
\end{eqnarray}
This extends the Lagrangian given in Ref.~\cite{Rubakov:2013kaa} to include
the Horndeski functions $G_4$ and $G_5$.
The Lagrangian~(\ref{original-Lag}) and the DBI conformal galileon theory
are included in the general framework defined by~(\ref{gen-Horn})
as specific cases.

In this paper, we further generalize~(\ref{gen-Horn})
and consider
\begin{eqnarray}
&&
G_2=e^{2(\alpha+1)\lambda\phi}g_2(Y),
\quad
G_3=e^{2\alpha\lambda\phi} g_3(Y),
\nonumber\\&&
G_4=\frac{M^2_{\rm Pl}}{2}+e^{2\alpha\lambda\phi}g_4(Y),
\quad
G_5=e^{-2\lambda\phi}g_5(Y),\label{gen:Lag}
\end{eqnarray}
where $\alpha$ ($>0$) is a new dimensionless parameter.
The four functions, $g_2$, $g_3$, $g_4$, and $g_5$,
are arbitrary as long as several conditions presented in this section
and in Sec.~\ref{Sec:Perturbations} are satisfied.
We assume, however, that $g_4(0)=0$, so that $G_4\to \mpl^2/2$ as $Y\to 0$.
The Horndeski theory with~(\ref{gen:Lag}) admits
the following {\em generalized galilean genesis} solution:
\begin{eqnarray}
e^{\lambda\phi}\simeq \frac{1}{\lambda\sqrt{2Y_0}}\frac{1}{(-t)},
\quad
H\simeq\frac{h_0}{(-t)^{2\alpha+1}}
\quad (-\infty <t<0),\label{gen:back}
\end{eqnarray}
for large $|t|$,
where $Y_0$ and $h_0$ are positive constants.
We see that $Y\simeq Y_0$ for this background.
The parameter $\alpha$ in the Lagrangian
results in controlling the evolution of the Hubble rate.
The scale factor is given by
\begin{eqnarray}
a\simeq 1+\frac{1}{2\alpha}\frac{h_0}{(-t)^{2\alpha}},
\end{eqnarray}
and hence the solution describes the universe
that starts expanding from Minkowski in the asymptotic past,
similarly to the original galilean genesis solution
which corresponds to the case of $\alpha =1$.
The ``slow-expansion'' model considered in Ref.~\cite{Liu:2011ns}
is reproduced by taking the particular functions $g_i$ with $\alpha =2$.
We thus obtain a one-parameter family of the generalized genesis solutions
as an alternative to inflation.
Note that, although the evolution of the scale factor is very different from
quasi-de Sitter,
the universe in this scenario is also {\em accelerating}:
$\partial_t(a H) > 0$, and hence fluctuation modes will leave the horizon
during the genesis phase.

Substituting Eq.~(\ref{gen:back}) to the background equations~(\ref{app-eq1})--(\ref{app-eq3})
and picking up the dominant terms at large $|t|$,
we have
\begin{eqnarray}
&&
{\cal E}
\simeq e^{2(\alpha +1)\lambda\phi}\hat\rho(Y_0)\simeq 0,\label{gen:eq1}
\\
&&
{\cal P}
\simeq 2{\cal G}(Y_0)\dot H+e^{2(\alpha +1)\lambda\phi}\hat p(Y_0)\simeq 0,\label{gen:eq2}
\end{eqnarray}
where
\begin{eqnarray}
\hat\rho(Y)&:=&2Y g_2'-g_2-4\lambda Y\left(\alpha g_3-Yg_3'\right),
\\
\hat p(Y)&:=&g_2-4\alpha \lambda Yg_3 
\nonumber\\&&+8(2\alpha +1)\lambda^2 Y(\alpha g_4-Yg_4') ,
\\
{\cal G}(Y)&:=&M^2_{\rm Pl}-4\lambda Y\left(g_5+Yg_5'\right),
\end{eqnarray}
an overdot stands for differentiation with respect to $t$, and
a prime for differentiation with respect to $Y$. 
The constant
$Y_0$ is determined as a root of
\begin{eqnarray}
\hat\rho(Y_0)=0,\label{co1}
\end{eqnarray}
and then
$h_0$ is determined from Eq.~(\ref{gen:eq2}) as
\begin{eqnarray}
h_0= -\frac{1}{2(2\alpha+1)(2\lambda ^2Y_0)^{\alpha +1}}\frac{\hat p(Y_0)}{{\cal G}(Y_0)}.\label{h0determined}
\end{eqnarray}
As will be seen shortly, this background is stable for ${\cal G}(Y_0)>0$.
Therefore, the above NEC violating solution is possible provided that
\begin{eqnarray}
\hat p(Y_0) <0.\label{co2}
\end{eqnarray}

As will be demonstrated in the next section,
the generalized genesis solution will develop a singularity $H\to\infty$ at some $t=t_{\rm sing}$,
as in the original genesis model.
We therefore assume that the genesis phase is matched onto the standard radiation-dominated universe
before $t=t_{\rm sing}$,
ignoring for the moment the detail of the reheating process.
In conventional general relativity, matching two different phases
can be done by imposing that the Hubble parameter is continuous across the two phases.
However, the matching conditions are modified in general scalar-tensor theories
as second-derivatives of the metric and the scalar field are mixed in the field equations.
The modified matching condition~\cite{Nishi:2014} reads
\begin{eqnarray}
\mpl^2 H_{\rm rad}&=&
{\cal G}(Y_0)H-\frac{e^{(2\alpha +1)\lambda\phi}}{2}\int^{Y_0}_0\sqrt{2y}\,g_3'(y)\D y
\nonumber\\&& \quad
+2\lambda\dot\phi e^{2\alpha\lambda\phi}\left(\alpha g_4-Y_0g_4'\right),
\end{eqnarray}
and we require that the subsequent radiation-dominated universe is expanding: $H_{\rm rad}>0$.
This condition translates to
\begin{eqnarray}
-g_2-2\lambda Y_0 g_3+(2\alpha+1)\lambda\sqrt{Y_0}\int_0^{Y_0}\frac{g_3}{\sqrt{y}}\D y>0.
\end{eqnarray}
It is easy to see that in the case of $\alpha =1$
all the expressions presented above reproduce the previous results~\cite{Nishi:2014}.

Before closing this section, let us emphasize that
(generalized) galilean genesis has the Minkowski phase only in the asymptotic past.
The true Minkowski spacetime solution corresponds to the special case of $Y=0$,
i.e., $\phi=$ const.
The $Y=0$ solution is found only if $g_2(0)=0$.
One may wonder if the true Minkowski vacuum ($Y=0$) in our neighborhood
begins to expand to form a genesis universe ($Y=Y_0>0$).
This is forbidden because the two different stable solutions cannot be interpolated,
as argued in Ref.~\cite{Rubakov:2013kaa}. (See, however, Ref.~\cite{Elder:2013gya}.)

\section{Background evolution}

To see whether or not the generalized genesis solution
presented in the previous section is an attractor,
we trace the background evolution starting from generic initial conditions.

\subsection{Analytic argument}

\begin{figure}[tbp]
  \begin{center}
    \includegraphics[keepaspectratio=true,height=70mm]{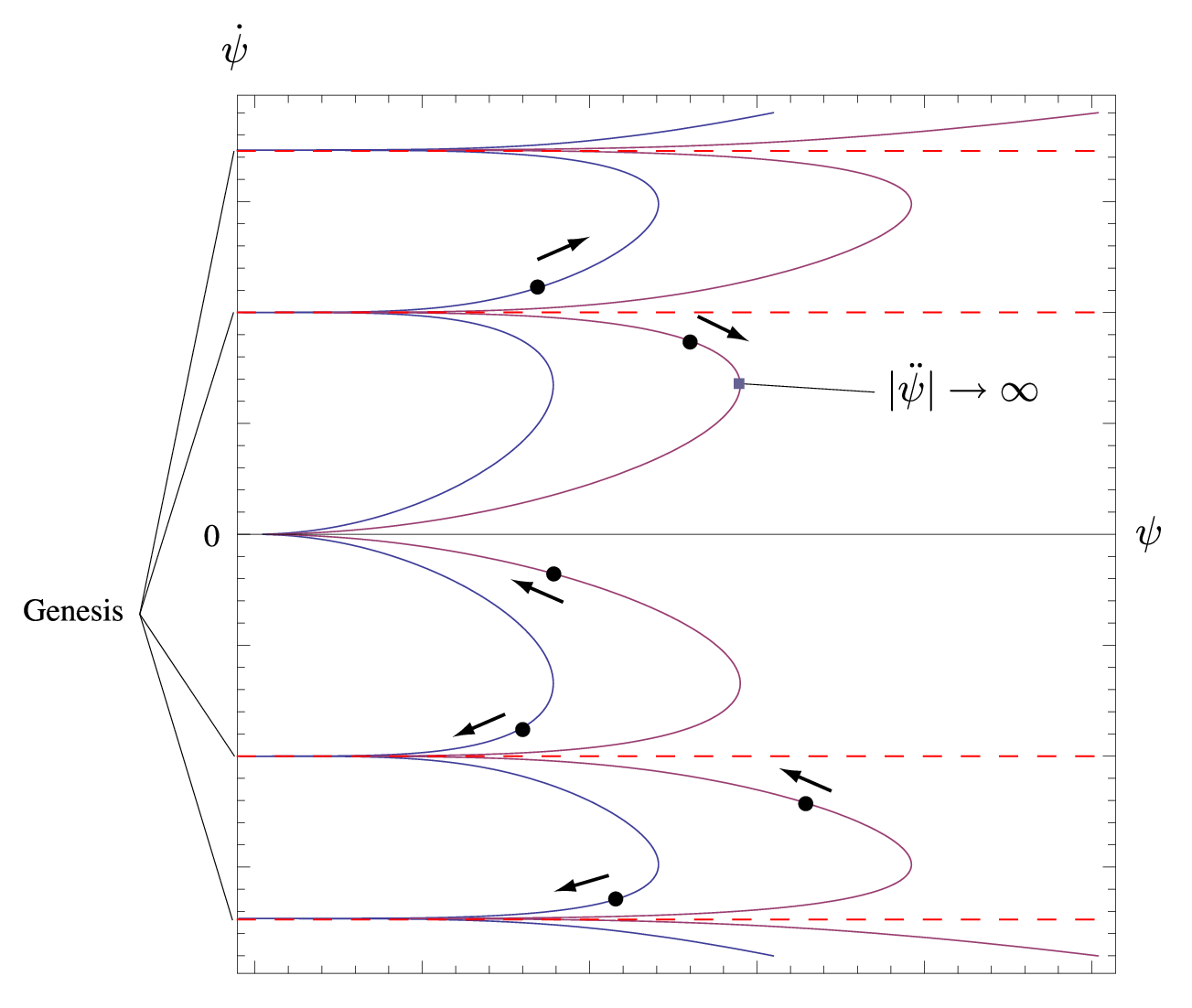}
  \end{center}
  \caption{
Examples of the curves defined by Eq.~(\ref{defcurve}).
Horizontal dashed lines correspond to the genesis solutions.
}%
  \label{fig:attractor.eps}
\end{figure}

Let us begin with a simplified discussion neglecting gravity, i.e.,
the effect of the cosmic expansion~\cite{Rubakov:2013kaa}.
It is convenient to introduce a new variable
\begin{eqnarray}
\psi:= e^{-\lambda \phi}\;(>0).
\end{eqnarray}
In terms of $\psi$ we have $Y=\dot\psi^2/(2\lambda^2)$.
For any homogeneous solutions the scalar-field equation of motion~(\ref{seomgeneral})
with the functions~(\ref{gen:Lag})
can be written as
\begin{eqnarray}
\frac{\D}{\D t}\left[\psi^{-2(\alpha+1)}\hat\rho(Y)\right]=0.
\end{eqnarray}
Integrating this, we obtain
\begin{eqnarray}
\hat\rho(Y)=C\psi^{2(\alpha +1)},\label{defcurve}
\end{eqnarray}
where $C$ is an integration constant.
Equation~(\ref{defcurve}) defines a curve in the $(\psi, \dot\psi)$ space for each $C$,
as shown in Fig.~\ref{fig:attractor.eps}.
With an initial condition $(\psi_{\rm i},\dot\psi_{\rm i})$ away from
the genesis solution, the integration constant is determined as
$C=\psi_{\rm i}^{-2(\alpha +1)}\hat\rho(\dot\psi_{\rm i}^2/2\lambda^2)$.
If $\dot\psi<0$ initially, the scalar field rolls along the curve
toward $\psi \to 0$, i.e., $\hat\rho \to 0$.
Hence, this solution approach to one of the genesis solutions which are denoted as horizontal lines
($\dot\psi =$ const) in the $(\psi, \dot\psi)$ plane.
If $\dot\psi>0$ initially, the scalar field rolls the opposite way along the curve
and goes further away from the genesis solutions.
This is the time reversal of the $\dot\psi <0$ solutions.

The above analytic argument implies that
the genesis solution is the attractor for initial conditions such that
$\dot\psi <0\;(\Leftrightarrow (e^{\lambda\phi}){\bf \dot{}}>0)$.
In the next subsection we perform numerical calculations
to show that this is basically true
even if one takes into account of the effect of gravity.
The numerical analysis also allows us to see the final fate of the genesis solutions
for which the effect of the cosmic expansion cannot be ignored.

\subsection{Full numerical analysis}

In the Horndeski theory with~(\ref{gen:Lag})
the Friedmann equation can be written as
\begin{eqnarray}
{\cal E}&=&e^{2(\alpha+1)\lambda\phi}\hat\rho(Y)
+6H\dot\phi e^{2\alpha\lambda\phi}c_1(Y)
\nonumber\\&&
-3H^2\left[c_2(Y)+e^{2\alpha\lambda\phi}d_2(Y)\right]
+2H^3\dot\phi e^{-2\lambda\phi}c_3(Y)
\nonumber\\
&=&0,\label{general-FR}
\end{eqnarray}
where
\begin{eqnarray}
c_1&=&Yg_3'-2\alpha\lambda g_4+2(3-2\alpha)\lambda Yg_4'+4\lambda Y^2g_4'',
\\
c_2&=&\mpl^2-12\lambda Yg_5-28\lambda Y^2g_5'-8\lambda Y^3g_5'',
\\
c_3&=&5Yg_5'+2Y^2g_5'',
\\
d_2&=&2g_4-8Yg_4'-8Y^2g_4''.
\end{eqnarray}
Equation~(\ref{general-FR}) is exact and
hence can be used even if the background evolution
is away from the genesis solution.
Similarly, one can substitute Eq.~(\ref{gen:Lag})
to the evolution equation ${\cal P}=0$ and the scalar-field equation of motion
to write straightforwardly the exact equations for the background.
The resultant equations are integrated numerically,
giving the background evolution starting from generic initial conditions.

Given the initial conditions $(\phi(t_0), \dot\phi(t_0))$,
the initial value for $H$ is determined from
the Friedmann equation~(\ref{general-FR}). Therefore,
the initial values $(\phi(t_0), \dot\phi(t_0))$ must be
chosen in such a way that Eq.~(\ref{general-FR})
admits a real root $H$.
Equation~(\ref{general-FR}) is quadratic in $H$ if $g_5=0$
and cubic if $g_5\neq 0$.
In both cases, the discriminant ${\cal D}$
for $e^{\lambda\phi}\ll 1$
is given by
\begin{eqnarray}
{\cal D}=e^{2(\alpha +1)\lambda\phi}c_2(Y)\hat\rho(Y)+{\cal O}(e^{2(2\alpha +1)\lambda\phi}).
\end{eqnarray}
In the $g_5=0$ case, the initial data $(\phi(t_0), \dot\phi(t_0))$
must lie in the region where ${\cal D}\ge 0$ is satisfied.
In the $g_5\neq 0$ case, the Friedmann equation has at least one real root
for any $(\phi(t_0), \dot\phi(t_0))$.

\begin{figure}[tbp]
  \begin{center}
    \includegraphics[keepaspectratio=true,height=70mm]{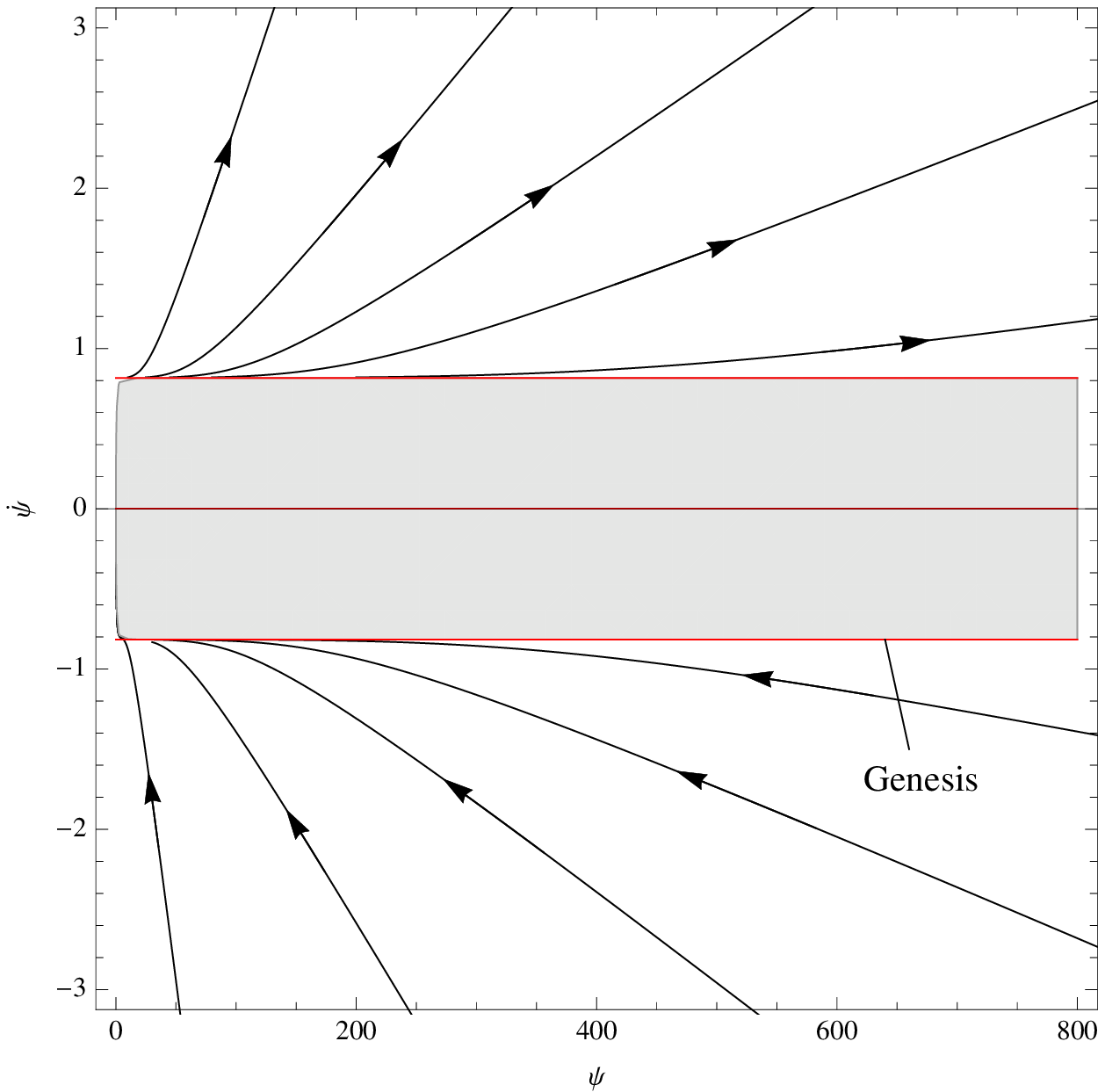}
  \end{center}
  \caption{Numerical results of the background evolution for
  the model with $g_2=-Y+Y^2$, $g_3=Y$, and $g_4=g_5=0$.
  The parameters are given by $\mpl =1$, $\lambda=1$, and $\alpha =1$.
}%
  \label{fig:attractorg2g3.eps}
\end{figure}
\begin{figure}[tbp]
  \begin{center}
    \includegraphics[keepaspectratio=true,height=70mm]{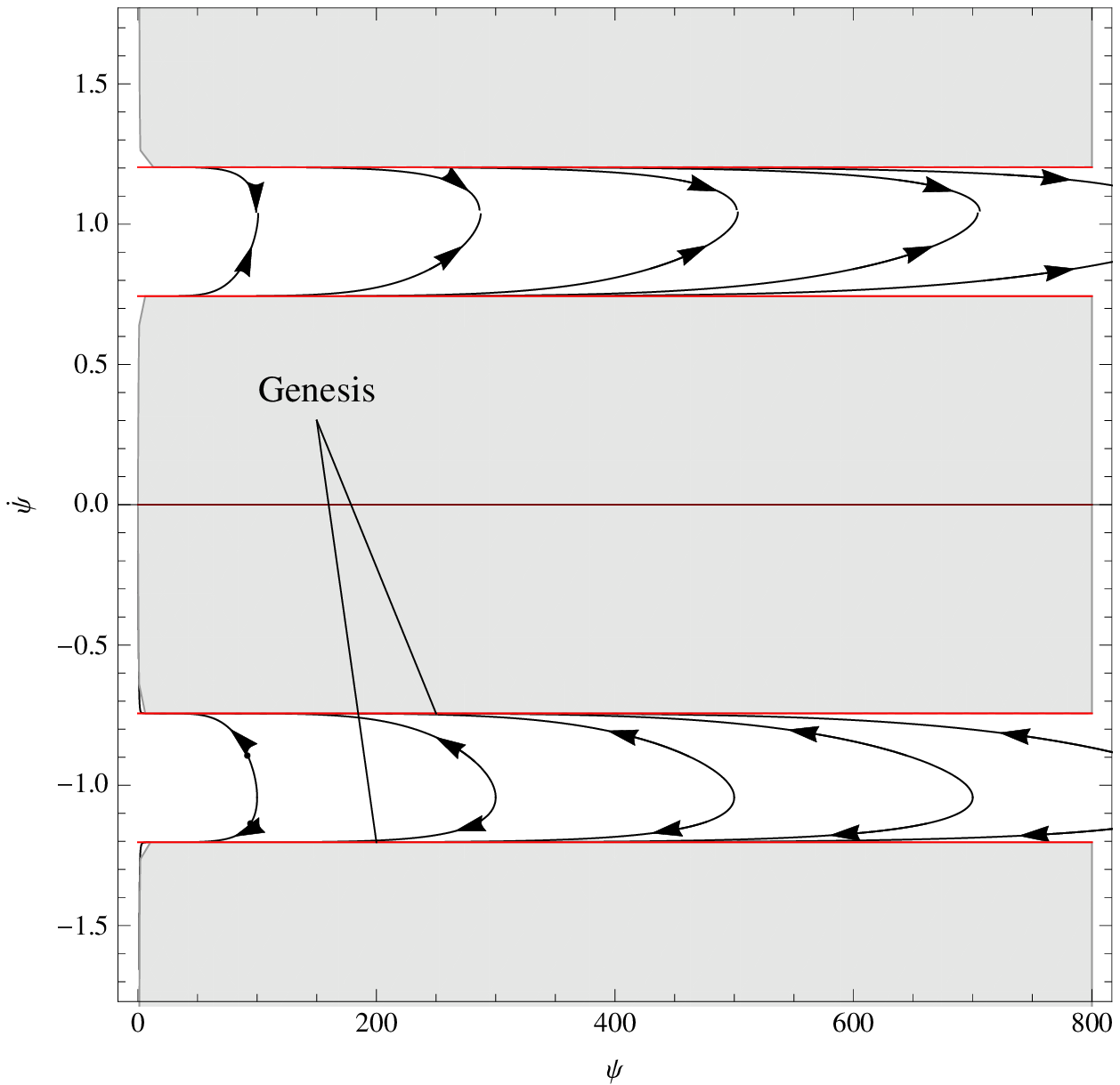}
  \end{center}
  \caption{Numerical results of the background evolution for
the model $g_2=-Y+3Y^2-Y^3$, $g_3=Y$, and $g_4=g_5=0$.
The parameters are given by $\mpl =1$, $\lambda=1$, and $\alpha =2$.
}%
  \label{fig:attractorg2g3-2.eps}
\end{figure}
\begin{figure}[tbp]
  \begin{center}
    \includegraphics[keepaspectratio=true,height=70mm]{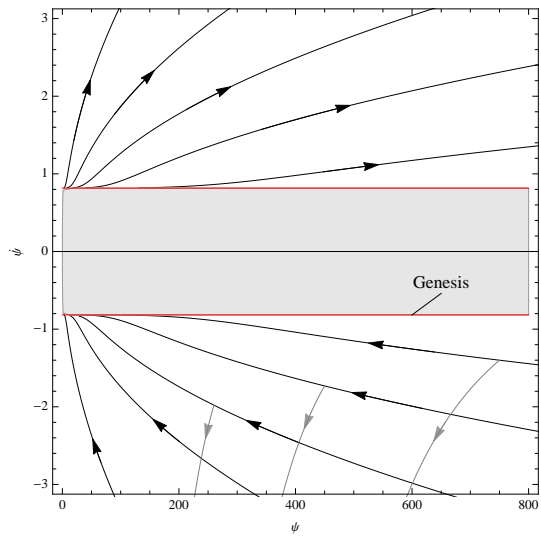}
  \end{center}
  \caption{Numerical results of the background evolution for
$g_2=-Y+Y^2$, $g_3=Y$, $g_4=0$, and $g_5=-Y$.
The parameters are given by $\mpl =1$, $\lambda=1$, and $\alpha =1$.
}%
  \label{fig:attractorg5.eps}
\end{figure}

Concrete numerical examples are presented in Figs.~\ref{fig:attractorg2g3.eps}--\ref{fig:attractorg5.eps}.
In Figs.~\ref{fig:attractorg2g3.eps} and~\ref{fig:attractorg2g3-2.eps}
we show the cases where the Friedmann equation is quadratic in $H$.
The shaded regions (${\cal D}<0$) cannot be accessed because $H$ would be imaginary there.
In Fig.~\ref{fig:attractorg2g3.eps} we have one genesis solution,
while we have two in Fig.~\ref{fig:attractorg2g3-2.eps}.
In both cases, generalized galilean genesis is the attractor for $\dot\psi <0$.
At late times where $\psi\ll 1$, the numerical solutions are no longer
approximated by Eq.~(\ref{gen:back}),
and within a finite time the Hubble rate $H$ diverges.
As the Friedmann equation is quadratic in $H$,
we have two branches of the solutions, one of which may be contracting initially ($H<0$).
The cosmological evolution nevertheless approaches the same genesis solution
and the trajectories in the $(\psi, \dot\psi)$ space are almost indistinguishable.

The behavior of the models with $g_5\neq 0$ is more complicated, as illustrated
in Fig.~\ref{fig:attractorg5.eps}.
In the white region, we have ${\cal D}>0$ and so there are three possible choices
for the initial value of $H$.
Two of the three branches converge to the genesis solution
similarly to the $g_5=0$ case, as shown as the black lines in Fig.~\ref{fig:attractorg5.eps}.
Also in this case we find $H\to \infty$ within a finite time.
However, the remaining one branch never converges to the genesis solution.
The corresponding examples are shown as the gray lines in Fig.~\ref{fig:attractorg5.eps}.
In the shaded region, we have ${\cal D}<0$ and there is only one possible initial value for $H$
at each point,
which corresponds to the latter branch.
Therefore, the generalized galilean genesis solution can be a dynamical attractor
for the initial data in the white (${\cal D}>0$) region.

We thus conclude that the galilean genesis solution is the attractor
provided that $\dot\psi <0\;(\Leftrightarrow (e^{\lambda\phi}){\bf \dot{}}>0)$ initially,
though the situation in the presence of $g_5$ is involved.
In the inflationary scenario, usually it does not matter which direction the scalar field rolls initially,
but the universe must be expanding initially.
In contrast to the case of inflation, the galilean genesis scenario
allows both for expanding and contracting universes at the initial moment,
while the scalar field must roll in a particular direction initially.
As far as we have investigated numerically,
all the solutions develop a singularity $H\to\infty$ at some time $t=t_{\rm sing}$ in the future.
In passing, we have checked that the numerical examples in Figs.~\ref{fig:attractorg2g3.eps} and~\ref{fig:attractorg5.eps} 
satisfy the stability conditions presented in the next section.

\subsection{Spatial curvature}

We have so far neglected the spatial curvature.
In this subsection, let us justify this assumption
by showing that the spatial curvature does not interfere with
the evolution of the genesis background.
We will use the cosmological background equations
with the spatial curvature $K$ in the Horndeski theory,
which are summarized in the Appendix~\ref{App:K}.

Let us take an initial condition such that
$H$ is sufficiently small in the equation of motion for $\phi$
and $\left(e^{\lambda\phi}\right){\bf \dot{}}>0$.
Then, in a universe with $K\neq 0$
the equation of motion for $\phi$ can be written as
\begin{eqnarray}
\frac{\D}{\D t}\left[e^{2(\alpha +1)\lambda \phi}\hat\rho(Y)
-e^{2\alpha\lambda\phi}\frac{{\cal K}_4(Y)}{a^2} -\frac{{\cal K}_5(Y)}{a^2}\right]
=0,\label{seomwk}
\end{eqnarray}
where
\begin{eqnarray}
{\cal K}_4(Y)&:=&6\left(g_4-2Yg_4'\right)K,
\\
{\cal K}_5(Y)&:=&-12\lambda Y\left(g_5+Yg_5'\right)K.
\end{eqnarray}
Even if $e^{2(\alpha +1)\lambda \phi}\hat\rho \sim e^{2\alpha\lambda\phi}{\cal K}_4, {\cal K}_5$
at the initial moment, the curvature terms become smaller
relative to the $\hat\rho$ term as the scalar field rolls.
Thus, we have the same attractor solution $Y=Y_0$ satisfying $\hat\rho(Y_0)=0$,
i.e., $e^{\lambda\phi}\sim (-t)^{-1}$.
Along this attractor, the evolution equation reads
\begin{eqnarray}
2{\cal G}(Y_0)\dot H + e^{2(\alpha+1)\lambda\phi}\hat p(Y_0)
+\left[\mpl^2+4\lambda Y_0g_5(Y_0)\right]\frac{K}{a^2}
\simeq 0,
\nonumber\\ \label{Peqwk}
\end{eqnarray}
where we assumed that $\dot H\gg H^2$.
Equation~(\ref{Peqwk}) implies that
the curvature term becomes subdominant
as the scalar field rolls, and as a result $\dot H$ is determined
by the $\hat p$ term, recovering the evolution of the genesis background.
Thus, the flatness problem is resolved in the genesis model.

\subsection{Anisotropy}

In conventional cosmology, an initial anisotropy is wiped out during inflation~\cite{Wald:1983ky}.
However, in alternative scenarios such as bouncing cosmology,
it is often problematic that the initial anisotropy grows
in a contracting phase~\cite{shearprob} (see however \cite{Cai:2013vm}).
In this subsection we will show that
adding the initial anisotropy on the generalized galilean genesis solution
does not destabilize the background evolution.

We consider the Kasner metric
\begin{eqnarray}
\D s^2=-\D t^2+a^2\left[e^{2\theta_1(t)}\D x^2+e^{2\theta_2(t)}\D y^2+e^{2\theta_3(t)}\D z^2\right],
\end{eqnarray}
where it is convenient to write
\begin{eqnarray}
\theta_1=\beta_++\sqrt{3}\beta_-,\quad \theta_2=\beta_+-\sqrt{3}\beta_-, \quad \theta_3=-2\beta_+ .
\end{eqnarray}
If the deviations from the genesis background are not large,
it follows from Eqs.~(\ref{aniso1}) and~(\ref{aniso2}) that
\begin{eqnarray}
\frac{\D}{\D t}\left[{\cal G}\dot\beta_+-2e^{-2\lambda\phi} \dot\phi Y_0g_5'\left(\dot\beta_+^2-\dot\beta_-^2\right)
\right]&=&0,
\\
\frac{\D}{\D t}\left[{\cal G}\dot\beta_-+4e^{-2\lambda\phi}\dot\phi  Y_0g_5'\dot\beta_+ \dot\beta_-
\right]&=&
0.
\end{eqnarray}
In the models with $g_5'=0$, this simply gives
\begin{eqnarray}
\dot\beta_+,\;\dot\beta_-\sim {\rm const},
\end{eqnarray}
so that the initial anisotropy dilutes as $\theta_i\sim (-t)$.
In the models with $g_5'\neq 0$,
we have the following possibilities:
\begin{eqnarray}
(\dot\beta_+,\dot\beta_-)=(0,0),(b,0),(-\frac{1}{2}b,\frac{\sqrt{3}}{2}b),(-\frac{1}{2}b,-\frac{\sqrt{3}}{2}b),
\end{eqnarray}
where
\begin{eqnarray}
b:=\frac{{\cal G}}{2e^{-2\lambda\phi}\dot\phi Y_0g_5'}\sim (-t)^{-1}.
\end{eqnarray}
In this case, for nonzero $\dot\beta_\pm$
the initial anisotropy can grow logarithmically: $\theta_i\sim \ln (-t)$.
However, this should be compared with $\ln a\sim (-t)^{-2\alpha}$; we see that
the logarithmic growth of $\theta_i$ does not spoil the genesis background.

\section{Primoridal perturbations}
\label{Sec:Perturbations}

Let us now study the behavior of primordial tensor and scalar perturbations
around the generalized genesis background
to obtain predictions of our scenario as well as
to impose stability conditions.
To do so, we utilize the general quadratic action for cosmological perturbations
in the Horndeski theory derived in Ref.~\cite{Kobayashi:2011nu}.

\subsection{Tensor perturbations}
The quadratic action
for tensor perturbations $h_{ij}$ in the genesis phase is given by
\begin{eqnarray}
S_h^{(2)}=\frac{1}{8}\int \D t\D^3x
\,a^3{\cal G}(Y_0)\left[\dot h^2_{ij}-\frac{c_t^2}{a^2}(\nabla h_{ij})^2\right],
\label{action_hij}
\end{eqnarray}
where
\begin{eqnarray}
c_t^2 =\frac{M^2_{\rm Pl}+4\lambda Y_0 g_5(Y_0)}{{\cal G}(Y_0)}
\end{eqnarray}
and note that $a\simeq 1$.
It can be seen that stability against tensor perturbations is assured if
\begin{eqnarray}
&&{\cal G}(Y_0)>0,\label{co3}\\
&&
M^2_{\rm Pl}+4\lambda Y_0 g_5(Y_0)>0,\label{co4}
\end{eqnarray}
are satisfied.

Since both ${\cal G}(Y_0)$ and $c_t^2$ are constant during the genesis phase,
the tensor perturbations are effectively living in Minkowski
without regard to $\alpha$ and the concrete form of $g_i(Y)$,
and consequently
amplification of quantum fluctuations does not occur as opposed to the case of
quasi-de Sitter inflation. This means that no detectable primordial gravitational waves are
generated from our generic class of the genesis models.

\subsection{Scalar perturbations}

The quadratic action for the curvature perturbation $\zeta$
in the unitary gauge
is given by
\begin{eqnarray}
S_\zeta^{(2)}=\int \D t\D^3x\,a^3{\cal G}_S
\left[  \dot\zeta^2 -\frac{c_s^2}{a^2} (\nabla \zeta)^2\right],\label{action-zeta}
\end{eqnarray}
where with some manipulation ${\cal G}_S$ and $c_s^2$
in the genesis phase are written as
\begin{eqnarray}
{\cal G}_S&=&2\left[
\frac{(2\alpha +1)\lambda\xi^2(Y_0)}{Y_0\xi'(Y_0)}
\right]^2\hat\rho'(Y_0)e^{-2\alpha\lambda\phi},\label{Gs}
\\
c_s^2&=&\frac{\xi'(Y_0)\hat p(Y_0)}{\xi(Y_0)\hat\rho'(Y_0)},\label{cs}
\end{eqnarray}
with
\begin{eqnarray}
\xi(Y):=-\frac{Y{\cal G}(Y)}{\hat p(Y)}.
\end{eqnarray}
Equations~(\ref{Gs}) and~(\ref{cs}) show that ${\cal G}_S\propto (-t)^{2\alpha}$
and $c_s^2=$ const.
It follows from Eqs.~(\ref{co2}) and~(\ref{co3}) that $\xi(Y_0)>0$.
We thus find that stability against scalar perturbations is guaranteed if
\begin{eqnarray}
&&
\hat\rho'(Y_0)>0,\label{co5}
\\
&&
\xi'(Y_0)<0,\label{co6}
\end{eqnarray}
are fulfilled.
We can choose the functional degrees of freedom so that this is possible.


Let us evaluate the power spectrum of $\zeta$.
To simplify the notation, it is convenient to write ${\cal G}_S={\cal A}(-t)^{2\alpha }$,
where ${\cal A}$ is a constant deduced from Eq.~(\ref{Gs}),
the value of which depends on the model, i.e., $\alpha$ and
the concrete form of $g_i(Y)$.
The equation of motion derived from the action~(\ref{action-zeta})
is given by
\begin{eqnarray}
\ddot\zeta_k+\frac{2\alpha}{t}\dot\zeta_k+c_s^2k^2\zeta_k=0 ,
\end{eqnarray}
where we moved to the Fourier space.
This equation can be solved to give
\begin{eqnarray}
\zeta_k =\frac{1}{2}\sqrt{\frac{\pi}{2{\cal A}(Y_0)}}(-t)^{\nu}H^{(1)}_{\nu}(-c_skt),
\quad\nu :=\frac{1}{2}-\alpha,
\label{eq_zeta}
\end{eqnarray}
where $H_\nu^{(1)}$ is the Hankel function of the first kind and
the positive frequency modes have been chosen.
On large scales, $|c_skt|\ll 1$, we have
\begin{eqnarray}
\zeta_k\simeq A_k+B_k(-t)^{1-2\alpha},\label{large-scale-sol}
\end{eqnarray}
where
\begin{eqnarray}
A_k&:=&-i2^{\nu-1}\sqrt{\frac{\pi}{2{\cal A}}}\frac{\Gamma(\nu)}{\pi}(c_sk)^{-\nu},
\\
B_k&:=& 2^{-\nu-1}\sqrt{\frac{\pi}{2{\cal A}}}
\left[
\frac{1}{\Gamma(\nu+1)}-\frac{i\cos(\pi\nu)\Gamma(-\nu)}{\pi}
\right]
\nonumber\\&& \times
(c_sk)^{\nu}.
\end{eqnarray}

If $0<\alpha <1/2$, the second term in Eq.~(\ref{large-scale-sol}) decays
as is common to usual cosmologies, leaving the constant mode at late times.
Thus, in this case the power spectrum is given by 
\begin{eqnarray}
{\cal P}_{\zeta}(k)=\frac{2^{2\nu-4}c_s^{-2\nu}\Gamma^2(\nu)}{\pi^3 {\cal A}}k^{3-2\nu},
\label{Psd1}
\end{eqnarray}
and the spectral index is found to be
\begin{eqnarray}
n_s=2\alpha+3,
\end{eqnarray}
yielding a blue spectrum
incompatible with observations.

The case of $\alpha >1/2$ is more subtle, because
the second term in Eq.~(\ref{large-scale-sol}) grows and dominates on large scales.
This is what happens in the original galilean genesis model ($\alpha =1$)~\cite{Creminelli:2010ba}.
To extract the late-time amplitude of $\zeta_k$, let us consider the following situation.
Suppose that the genesis phase terminates at $t=t_{\rm end}$
and is matched onto some other phase.
We assume that the scalar field is homogeneous on the $t=t_{\rm end}$ hypersurface.
In the subsequent phase,
the curvature perturbation on large scales may be written as
\begin{eqnarray}
\zeta_k=C_k-D_k\int_t^\infty\frac{\D t'}{a^3(t'){\cal G}_S(t')},\label{late-sol}
\end{eqnarray}
where we do not specify ${\cal G}_S(t)$ for $t>t_{\rm end}$,
but assume that Eq.~(\ref{late-sol}) gives the constant and decaying modes and hence the integral converges.
The late-time amplitude is given by $C_k$.
The matching conditions~\cite{Nishi:2014} imply that
$\zeta_k$ and ${\cal G}_S\dot\zeta_k$ are continuous across the two phases (cf.~\cite{Hwang:1991an}).
It is then straightforward to obtain
$C_k=A_k+B_k(-t_{\rm end})^{1-2\alpha}\left(1+{\cal I}\right)\simeq B_k(-t_{\rm end})^{1-2\alpha}\left(1+{\cal I}\right)$,
where
\begin{eqnarray}
{\cal I}:=(2\alpha -1)\int_{t_{\rm end}}^\infty
\frac{a^3(t_{\rm end}){\cal A}(-t_{\rm end})^{2\alpha}}{a^3(t'){\cal G}_S(t')}\frac{\D t'}{|t_{\rm end}|}
\end{eqnarray}
is independent of $k$. We may thus use the estimate
\begin{eqnarray}
{\cal P}_\zeta(k) \sim {\cal C}\times \left.{\cal P}_\zeta(k)\right|_{t=t_{\rm end}},
\end{eqnarray}
with ${\cal C}$ being some $k$-independent factor.
The power spectrum evaluated at the end of the genesis phase is given by
\begin{eqnarray}
\left.{\cal P}_\zeta(k)\right|_{t=t_{\rm end}}
=\frac{2^{-2\nu-4}c_s^{2\nu}\Gamma^2(-\nu)}{\pi^3{\cal A}}\left|t_{\rm end}\right|^{4\nu}
k^{3+2\nu},
\end{eqnarray}
so that
\begin{eqnarray}
n_s=5-2\alpha.
\end{eqnarray}
Although the overall amplitude depends on the details of the model construction,
the spectral index depends only on $\alpha$ and not on the concrete form of $g_i(Y)$.
We have an exactly scale-invariant spectrum for $\alpha =2$, and
this is in sharp contrast to the original galilean genesis model having $\alpha =1$,
which produces a blue-tilted spectrum of curvature perturbations.
A particular realization of $\alpha=2$
is found in Ref.~\cite{Liu:2011ns,Piao:2010bi}, where the same conclusion is reached.  
Taking $\alpha = 2.02$,
one can obtain the nearly scale-invariant, but slightly red-tilted, spectrum with $n_s\simeq 0.96$.

\section{Curvaton}

In the previous section we have seen that
the nearly scale-invariant spectrum for curvature perturbations
is possible only in the case of $\alpha\simeq 2$.
In the other cases we need to consider an alternative mechanism such as the curvaton
in order to obtain a scale-invariant spectrum.
In this section, we study slightly in more detail the curvaton
coupled to a conformal metric, the basic idea of which was
proposed earlier in Ref.~\cite{Creminelli:2010ba}.
A similar mechanism was proposed in Ref.~\cite{Wang:2012bq}.

To make a scale-invariant power spectrum,
we introduce a curvaton field $\sigma$ coupled to the conformal metric,
\begin{eqnarray}
\hat{g}_{\mu\nu}=e^{2\beta\lambda\phi}g_{\mu\nu},\label{confmet}
\end{eqnarray}
where $\beta$ is a constant parameter which is assumed to be close to unity, $\beta\simeq 1$.
Assuming the simplest potential, we consider the following action for $\sigma$:
\begin{eqnarray}
S_{\sigma}=\int\D^4x\sqrt{-\hat g}
\left[-\frac{1}{2}\hat{g}^{\mu\nu}\partial_{\mu}\sigma\partial_{\nu}\sigma-\frac{1}{2}m^2\sigma^2\right].
\end{eqnarray}
The conformal metric~(\ref{confmet}) implies that the effective scale factor
for the curvaton is $e^{\beta\lambda\phi}\sim (-t)^{-\beta}$ with $\beta\simeq 1$,
so that $\sigma$ lives effectively in a quasi-de Sitter spacetime.

The equations of motion for the homogeneous part $\sigma=\sigma_0(t)$ is given by
\begin{eqnarray}
\ddot\sigma_0+(2\beta\lambda\dot\phi+3H)\dot\sigma_0
+e^{2\beta\lambda\phi}m^2\sigma_0=0.
\label{eom_sigma}
\end{eqnarray}
On the genesis background, one can ignore $H\sim (-t)^{-(2\alpha +1)}$ relative to $\lambda\dot\phi\sim (-t)^{-1}$,
leading to
\begin{eqnarray}
\ddot\sigma_0-\frac{2\beta}{t}\dot\sigma_0+\frac{m^2}{[\lambda\sqrt{2 Y_0} (-t)]^{2\beta}}\sigma_0=0.
\end{eqnarray}
The effective Hubble rate for the curvaton is $\sim \lambda\sqrt{2Y_0}$.
For the ``light'' curvaton with
\begin{eqnarray}
m^2\ll\lambda^2 Y_0,
\end{eqnarray}
we thus have $\sigma_0\simeq$ const and the other independent solution decays quickly.

The energy density and pressure of $\sigma$ are given by
\begin{eqnarray}
\rho_{\sigma}&=& \frac{1}{2}e^{2\beta \lambda \phi}\dot\sigma_0^2 +\frac{1}{2}e^{4\beta\lambda\phi}m^2\sigma_0^2
\sim (-t)^{-4\beta},
\\
p_{\sigma}&=&\frac{1}{2}e^{2\beta \lambda \phi}\dot\sigma_0^2 -\frac{1}{2}e^{4\beta\lambda\phi}m^2\sigma_0^2
\sim (-t)^{-4\beta}.
\end{eqnarray}
Equations~(\ref{gen:eq1}) and~(\ref{gen:eq2})
imply that the dominant part of the cosmological background equations grows as $\sim (-t)^{-2(\alpha +1)}$.
Thus, in order for the (initially subdominant)
curvaton not to spoil the genesis background as time proceeds,
we require that
\begin{eqnarray}
\alpha + 1 \ge 2\beta.\label{condition-curv}
\end{eqnarray}

The fluctuation of the curvaton, $\delta\sigma(t, {\mathbf x})$,
obeys
\begin{eqnarray}
\ddot{\delta\sigma}
-\frac{2\beta}{t}\dot{\delta\sigma}-\nabla^2\delta\sigma+
\frac{m^2}{[\lambda\sqrt{2 Y_0} (-t)]^{2\beta}}\delta\sigma=0.
\label{eom_dsigma}
\end{eqnarray}
Neglecting the mass term,
this can be solved in the Fourier space to give
\begin{eqnarray}
\delta\sigma_k=\frac{\sqrt{\pi}}{2}\left(\lambda\sqrt{2Y_0}\right)^\beta (-t)^{\beta+1/2}
H_{\beta+1/2}^{(1)}(-kt),
\end{eqnarray}
where the positive frequency modes have been chosen.
Thus, the power spectrum of the curvaton fluctuations is
\begin{eqnarray}
{\cal P}_{\delta\sigma}(k)=
\frac{2^{3\beta-2}\lambda^{2\beta}Y_0^{\beta}\Gamma^2(\beta+1/2)}{\pi^3}k^{2-2\beta},
\end{eqnarray}
and we find
\begin{eqnarray}
n_s=3-2\beta.
\end{eqnarray}
In the case of $\beta =1$, the effective scale factor for the curvaton
is that of exact de Sitter, and hence the power spectrum is exactly scale-invariant, as is expected.
Taking $\beta = 1.04$ we obtain $n_s=0.96$.
The curvaton fluctuations can be converted into adiabatic ones after
the genesis phase, where $\sigma$ behaves as a conventional scalar field
in a true expanding universe, in the same way as the usual curvaton field
in the inflationary scenarios.
Note, however, that due to the restriction~(\ref{condition-curv})
the present curvaton mechanism works only for the models with $\alpha\ge 2-n_s \,(> 1)$.

\section{Conclusions}

In this paper, we have
extended the galilean genesis models~\cite{Creminelli:2010ba,Creminelli:2012my,Hinterbichler:2012fr,Hinterbichler:2012yn} and
constructed a generic Lagrangian from the Horndeski theory
that admits the {\em generalized galilean genesis} solution.
In generalized galilean genesis,
the universe starts expanding from Minkowski in a singularity free manner
with the increasing Hubble rate $H\sim (-t)^{-(2\alpha+1)}$,
where $\alpha\;(>0)$ is a new constant parameter in the Lagrangian.
We have investigated the background evolution and
shown that the generalized galilean genesis solution
is the attractor for a wide range of initial conditions.
In particular, we have seen that the spatial curvature and an initial anisotropy
do not hinder the evolution of the genesis phase.

We have then studied the primordial perturbations from the generalized galilean genesis models.
From the quadratic actions for cosmological perturbations we
have imposed several stability conditions on the functions in our generic Lagrangian.
In contrast to the case of quasi-de Sitter inflation,
tensor fluctuations are not amplified in the genesis phase
in all the galilean genesis models we have constructed,
and hence no detectable primordial gravitational waves are expected.
The evolution of the curvature perturbation $\zeta$ depends on the parameter $\alpha$ and
has turned out to be more interesting.
In the case of $\alpha >1/2$, $\zeta$ grows on large scales, as in
the original galilean genesis model ($\alpha=1$)~\cite{Creminelli:2010ba}.
The tilt of the power spectrum at the end of the genesis phase
is given by $n_s=5-2\alpha$, irrespective of the other details of the model.
Thus, we have a slightly red-tilted spectrum for $\alpha \gtrsim 2$.
In the case of $\alpha<1/2$, the constant mode dominates on large scales
as in conventional cosmology. In this case,
the power spectrum has been shown to be always blue-tilted.
We have also discussed the possibility of the curvaton mechanism
in the generalized galilean genesis scenario.

We have ignored the reheating process in our scenario.
It would be interesting to
explore how the universe reheats and how matter is created
at the end of generalized galilean genesis.
Since the Lagrangian defined by~(\ref{gen:Lag})
excludes a cosmological constant, it is not clear how
the genesis phase is connected finally to the late-time universe described by
the $\Lambda$CDM model. These are the open questions.

\acknowledgments
We would like to thank A. Vikman for useful comments.
This work was supported in part by JSPS Grant-in-Aid for Young
Scientists (B) No.~24740161 (T.K.). 


\appendix

\begin{widetext}
\section{Cosmological background equations in the Horndeski theory}
\label{App:K}

The cosmological background equations
in the Horndeski theory are given by~\cite{Kobayashi:2011nu}
\begin{eqnarray}
{\cal E}=0,\quad{\cal P}=0,\label{app-eq1}
\end{eqnarray}
where
\begin{eqnarray}
{\cal E}&:=&2XG_{2X}-G_2
+6X\dot\phi HG_{3X}-2XG_{3\phi}
-6H^2G_4+24H^2X(G_{4X}+XG_{4XX})
-12HX\dot\phi G_{4\phi X}-6H\dot\phi G_{4\phi }
\nonumber\\
&&+2H^3X\dot\phi\left(5G_{5X}+2XG_{5XX}\right)
-6H^2X\left(3G_{5\phi}+2XG_{5\phi X}\right),
\label{app-eq2}
\\
{\cal P}&:=&G_2
-2X\left(G_{3\phi}+\ddot\phi G_{3X} \right)
+2\left(3H^2+2\dot H\right) G_4
-12 H^2 XG_{4X}-4H\dot X G_{4X}
-8\dot HXG_{4X}-8HX\dot X G_{4XX}
\nonumber\\&&
+2\left(\ddot\phi+2H\dot\phi\right) G_{4\phi}
+4XG_{4\phi\phi}
+4X\left(\ddot\phi-2H\dot\phi\right) G_{4\phi X}
-2X\left(2H^3\dot\phi+2H\dot H\dot\phi+3H^2\ddot\phi\right)G_{5X}
\nonumber\\&&
-4H^2X^2\ddot\phi G_{5XX}
+4HX\left(\dot X-HX\right)G_{5\phi X}
+2\left[2\left(HX\right){\bf \dot{}}+3H^2X\right]G_{5\phi}
+4HX\dot\phi G_{5\phi\phi}.
\label{app-eq3}
\end{eqnarray}
The equation of motion for $\phi$ takes the form
\begin{eqnarray}
\frac{1}{a^3}\frac{\D}{\D t}\left(a^3J \right)&=&
P_\phi,\label{seomgeneral}
\end{eqnarray}
where
\begin{eqnarray}
J&:=&\dot\phi G_{2X}+6HXG_{3X}-2\dot\phi G_{3\phi}
+6H^2\dot\phi\left(G_{4X}+2XG_{4XX}\right)-12HXG_{4\phi X}
\nonumber\\&&
+2H^3X\left(3G_{5X}+2XG_{5XX}\right)
-6H^2\dot\phi\left(G_{5\phi}+XG_{5\phi X}\right), 
\end{eqnarray}
and
\begin{eqnarray}
P_\phi&:=&G_{2\phi}-2X\left(G_{3\phi\phi}+\ddot\phi G_{3\phi X}\right)
+6\left(2H^2+\dot H\right)G_{4\phi}
\nonumber\\&&
+6H\left(\dot X+2HX\right)G_{4\phi X}
-6H^2XG_{5\phi\phi}+2H^3X\dot\phi G_{5\phi X}.
\end{eqnarray}

The above equations are for the spatially flat background.
In the open ($K=-1$) and closed ($K=1$) cases, the corresponding equations are given by
\begin{eqnarray}
{\cal E} - \frac{3{\cal G}_TK}{a^2}=0,
\quad
{\cal P}+\frac{{\cal F}_T K}{a^2}=0,\label{FRW-K}
\end{eqnarray}
and
\begin{eqnarray}
&&
\frac{1}{a^3}\frac{\D}{\D t}\left\{a^3\left[
J+\frac{6K}{a^2}(\dot\phi G_{4X}+HXG_{5X}-\dot\phi G_{5\phi})
\right]\right\}
= P_\phi + \frac{3K}{a^2}\frac{\partial{\cal F}_T}{\partial \phi},
\end{eqnarray}
where
\begin{eqnarray}
{\cal G}_T&:=&2\left[
G_4-2XG_{4X}-X\left(H\dot\phi G_{5X} -G_{5\phi}\right)
\right],
\\
{\cal F}_T&:=&2\left[
G_4-X\left(\ddot\phi G_{5X}+G_{5\phi}\right)
\right].
\end{eqnarray}
The cosmological background equations for the open and closed models
are derived for the first time in this paper.

\section{Anisotropic Kasner universe in the Horndeski theory}
\label{App:Aniso}

We derive the basic equations governing the evolution of
an anisotropic Kasner universe in the Horndeksi theory.
We consider the following metric:
\begin{eqnarray}
\D s^2=-N^2\D t^2+a^2\left[
e^{2(\beta_++\sqrt{3}\beta_-)}\D x^2
+e^{2(\beta_+-\sqrt{3}\beta_-)}\D y^2
+e^{-4\beta_+}\D z^2
\right].
\end{eqnarray}
Substituting this to the Horndeski action, we obtain
\begin{eqnarray}
S=S_{\rm iso} +S_{\rm aniso},\label{action-iso+aniso}
\end{eqnarray}
where $S_{\rm iso}$ is identical to the action for the homogeneous and isotropic metric
and $S_{\rm aniso}$ is given by
\begin{eqnarray}
S_{\rm aniso}=
\int\D t\D^3x\left[\frac{6a^3}{N}
\left(
G_4-2XG_{4X}-\frac{H\dot\phi}{N^2}XG_{5X}+XG_{5\phi}
\right)
\left(\dot\beta_+^2+\dot\beta_-^2\right)
-\frac{4a^3}{N^3}X\dot\phi G_{5X}
\left(\dot\beta_+^3-3\dot\beta_+\dot\beta_-^2\right)\right].
\end{eqnarray}
Note that $X$ here should be understood as $X=\dot\phi^2/2N^2$.

Varying the above action with respect to $\beta_{\pm}$ and setting $N=1$, we obtain
\begin{eqnarray}
\frac{\D}{\D t}\left\{a^3\left[{\cal G}_T
\dot\beta_+
-2X\dot\phi G_{5X}\left(\dot\beta_+^2-\dot\beta_-^2\right)
\right]\right\}&=&0,\label{aniso1}
\\
\frac{\D}{\D t}\left\{a^3\left[{\cal G}_T
\dot\beta_-+4X\dot\phi G_{5X}\dot\beta_+\dot\beta_-
\right]\right\}&=&0.\label{aniso2}
\end{eqnarray}
Varying the action~(\ref{action-iso+aniso}) with respect to $N$, $a$, and $\phi$,
one can also derive the cosmological evolution equations
with shear contributions.

\end{widetext}



\end{document}